\def\spose#1{\hbox to 0pt{#1\hss}}

\def\multleft#1{\hbox to size{\vbox {\halign {\lft{##}\cr #1}}\hfill}\par}
\def\multright#1{\hbox to size{\vbox {\halign {\rt{##}\cr #1}}\hfill}\par}

\def\today{\ifcase\month\or January\or February\or March\or April\or May\or
      June\or July\or August\or September\or October\or November\or December\fi
      \space\number\day, \number\year}
\def\s{\hbox{\phantom{5}}}	

\def\cm{{\rm\thinspace cm}}

\def\erg{{\rm\thinspace erg}}

\def\K{{\rm\thinspace K}}
\def\keV{{\rm\thinspace keV}}
\def\km{{\rm\thinspace km}}
\def\kpc{{\rm\thinspace kpc}}

\def\m{{\rm\thinspace m}}
\def\Mpc{{\rm\thinspace Mpc}}
\def\Msun{\hbox{$\rm\thinspace M_{\odot}$}}
\def\pc{{\rm\thinspace pc}}

\def\s{{\rm\thinspace s}}

\def\yr{{\rm\thinspace yr}}

\def\ergpcmsqps{\hbox{$\erg\cm^{-2}\s^{-1}\,$}}

\def\ergps{\hbox{$\erg\s^{-1}\,$}}

\def\kmps{\hbox{$\km\s^{-1}\,$}}

\def\MsunMpc{\hbox{$\Msun\pc^{-3}\,$}}

\def\pcm{\hbox{$\cm^{-3}\,$}}

\def\ps{\hbox{$\s^{-1}\,$}}
\def\psqcm{\hbox{$\cm^{-2}\,$}}

\def\kmpspMpc{\hbox{$\kmps\Mpc^{-1}$}}

\def\H2{\hbox{H$_{2}$}}

\documentclass[usegraphicx]{mn2e}

\usepackage{times}
\usepackage{amssymb}
\include{defn}

\begin{document}
\hsize=6truein

\title[The nature of the molecular gas system in the core of NGC~1275]{The nature of the molecular gas system in the core of NGC~1275}
\author[R.J.~Wilman et al.]
{\parbox[]{6.in} {R.J.~Wilman$^{1}$, A.C.~Edge$^{1}$ and R.M.~Johnstone$^{2}$ \\ \\
\footnotesize
1. Department of Physics, University of Durham, South Road, Durham, DH1 3LE. \\
2. Institute of Astronomy, Madingley Road, Cambridge, CB3 0HA. \\ }}
\maketitle

\begin{abstract}
We present near-infrared integral field spectroscopy of the central kiloparsec of NGC~1275 at the 
heart of the Perseus cluster of galaxies, obtained with the UIST IFU on UKIRT. The nuclear 
ro-vibrational \H2~emission is spatially resolved and is likely to originate approximately 50\pc~from the 
active nucleus. The Pa$\alpha$ emission is, by contrast, spatially unresolved. The requirements for 
thermal excitation of the \H2~by nuclear X-radiation, its kinematics 
on sub-arcsec~scales, and its stability against self-gravity, together suggest that the observed \H2~is part of clumpy 
disk rotating about the radio-jet axis. The sharp jump in the \H2~velocity across the nucleus implies a black hole 
mass of $3.4 \times 10^{8}$\Msun, with a systematic error of $\pm 0.18$~dex due to the uncertainty in the radio jet inclination. This agrees well with the value implied by the empirical correlation between black 
hole mass and stellar velocity dispersion for nearby elliptical galaxies, and is $\sim 100$ times the stellar mass in this region. 
\end{abstract}


\section{INTRODUCTION}
For several years, NGC~1275 at the centre of the Perseus cluster held the unique distinction
of being the only galaxy in a cooling flow cluster for which CO emission from molecular gas had been 
detected (Inoue et al.~1996; Bridges \& Irwin~1998 and references therein). The fact that it is also a well-known 
galaxy merger (as well as having been classified a Seyfert 1, a BL Lac and an FR I radio source) served to 
obscure any causal connection between the CO and the cooling flow, so the failure to detect molecular gas in other 
such systems led to a heated debate over the validity of the cooling flow model. The latter model proposed 
that hot gas at the centres of relaxed clusters of galaxies was cooling out from the hot phase at 
rates of hundreds or thousands of solar masses per year (Fabian~1994). Recent developments on two
separate fronts have, however, helped to bridge the impasse. 

Firstly, X-ray data from Chandra and XMM-Newton show that cooling rates were vastly overestimated in the past 
(e.g. Schmidt, Allen \& Fabian~2001), and X-ray grating spectra also reveal a deficit of line emission from gas 
cooling below temperatures $\sim T_{\rm{virial}}/3$ (Peterson et al.~2003). Explanations for these findings 
include rapid mixing of hot and cold phases, inhomogeneously distributed metals in the ICM (Fabian et 
al.~2001,~2002), AGN heating by jets (Br\"{u}ggen \& Kaiser~2002) and sound waves (Fabian et al.~2003), thermal 
conduction (Voigt et al.~2002), and a significant relativistic cosmic ray component frozen into the thermal gas 
(Cen~2005). 

Secondly, using more sensitive receivers, Edge~(2001) has detected CO emission in 16 cooling flow central galaxies, 
consistent with $10^{9}-10^{11.5}$\Msun~of \H2~at 40\K~(see also Salom\'e \& Combes~2003). These are roughly the 
masses expected, given the revised cooling rates and likely ages. Interferometry shows further that the CO emission is localised within the central few arcsec of the cluster (Edge \& Frayer 2003; Salom\'e \& Combes~2004). The frequent 
occurrence of a much hotter molecular gas component in these systems has also been established, via near-infrared 
surveys of the ro-vibrational \H2~lines. Building on some earlier work (Elston \& Maloney~1992; Jaffe \& Bremer~1997; Falcke et al.~1998; Jaffe, Bremer \& van der Werf~2001), Edge et al.~(2002) 
performed an H+K band spectroscopic survey of 32 line-luminous central cluster galaxies and showed that the 
\H2~emission correlates in strength with the CO and H$\alpha$ emission. Analysis by Wilman et al.~(2002) showed that 
the lower-lying H$_{2}$ lines are 
thermally excited in dense gas ($n > 10^{5}$~cm$^{-3}$) at $T \sim 2000$K. The inferred gas pressure 
($nT$) thus exceeds $10^{8}$~cm$^{-3}$~K, some 2--3 orders of magnitude higher than that in either the 
diffuse X-ray or optical emission line gas. There must be a close physical connection between the H$_{2}$ and 
optical emission line clouds, since the H$_{2}$/H$\alpha$ ratio is observed to be constant over 2 orders of magnitude 
in H$\alpha$ luminosity. Narrow-band HST NICMOS imaging of three central cluster galaxies by Donahue et al.~(2000) 
had earlier shown that the hot H$_{2}$ and hydrogen recombination emission lines have similar morphologies.

Following on from our spectroscopic survey in Edge et al.~(2002), we have begun detailed studies of individual 
objects to analyse the relationship between the various ionized and molecular gas systems. Here we present H and 
K-band area spectroscopy of the centre of NGC~1275 using the UIST Integral Field 
Unit on the United Kingdom Infrared Telescope (UKIRT). From narrow-band HST observations of this target, 
Donahue et al.~(2000) found that the bulk of the ro-vibrational emission originates in the nuclear
source, with some diffuse emission lying up to a kiloparsec away. With the UIST IFU we can map this emission, infer its 
excitation state and explore its possible connection to the 1.2~\kpc~radius ring of CO-emission found by Inoue et 
al.~(1996). Area spectroscopy of NGC~1275 was published earlier by Krabbe et al.~(2000) but at significantly lower 
spatial and spectral resolution and non-simultaneously in the H and K-bands. 

The structure of the results section of this paper is as follows. Firstly, we analyse the emission line morphology in the central region and demonstrate that whilst the AGN continuum is (as expected) unresolved, the \H2~emission appears to be concentrated around $\sim 50$\pc~from the nucleus. Thereafter, we show that consideration of the excitation and kinematic properties of the \H2~are consistent with this interpretation, and suggest that the gas is in a disk-like structure spread over radii $\sim 35-70$\pc. We offer further support for the model by deriving a dynamical estimate of the mass of the central supermassive black hole, and find that it is consistent with the best non-dynamical estimates in the literature. We then discuss the possible connection between this \H2~disk and the CO emission on larger scales.

With $H_{\rm{0}}=70$~\kmpspMpc, the spatial scale in NGC~1275 (z=0.0176) is 358\pc~per arcsec.

\section{OBSERVATIONS, DATA REDUCTION AND ANALYSIS}
The observations were performed on the nights of 2003 September 24 and 26 using the UIST IFU on UKIRT. 
The IFU provides a $6.5 \times 3.4$~\arcsec~field of view and works on the image slicing principle, 
with the focal plane being covered by 14 `slits' which are aligned parallel to the long-axis of the field of view and 
reformatted into a single long slit at the entrance to the dispersion unit. The slits are 0.24\arcsec~wide and the 
pixel size along the slits is 0.12\arcsec. We performed observations with two separate grisms. The first observation 
was taken with the short-K grism spanning 2.05--2.25 \micron~with a spectral resolution of $\sim 3600$, and we 
observed the target with 500 second exposures for a total of 42 minutes on source (with the usual 
object-sky-sky-object nodding pattern). A second observation was obtained with the HK grism 
(resolution $\sim 900$), spanning 1.45--2.5\micron, with 240 second exposures for a total of 52 
minutes on source. The long-axis of the IFU was oriented in an east-west direction for both 
observations. The observations were taken under photometric conditions with atmospheric seeing 
in the range 0.3--0.4\arcsec, which is well-matched to the spatial sampling provided by UIST. 
For convenience, the two datasets will hereafter be referred to as the `short-K' and `HK' data, 
respectively.

Data reduction was performed offline using the UIST-specific recipes within the ORAC-DR software (Cavanagh et 
al. 2003) and comprised the usual steps of flat-fielding, sky-subtraction, wavelength calibration, division by 
an atmospheric standard and flux calibration. The resulting data cubes contain $14 \times 54$ spatial
elements for each of the 1024 spectral channels. They were cleaned of cosmetic defects such as hot 
pixels and cosmic rays by use of a median filtering process, whereby a two-dimensional kernel was 
passed through the cube and at each point pixels deviant from the median value by more than a certain 
threshold were replaced by the median. The filter was applied five times, during the course of which
the replacement threshold was gradually lowered from 5$\sigma$ to 2$\sigma$. Subsequent cube 
manipulation and spectral analysis were performed with a combination of IDL, IRAF and QDP/PGPLOT.

\section{RESULTS}

\subsection{Emission line morphologies}
\begin{figure*}
\includegraphics[width=0.84\textwidth,angle=0]{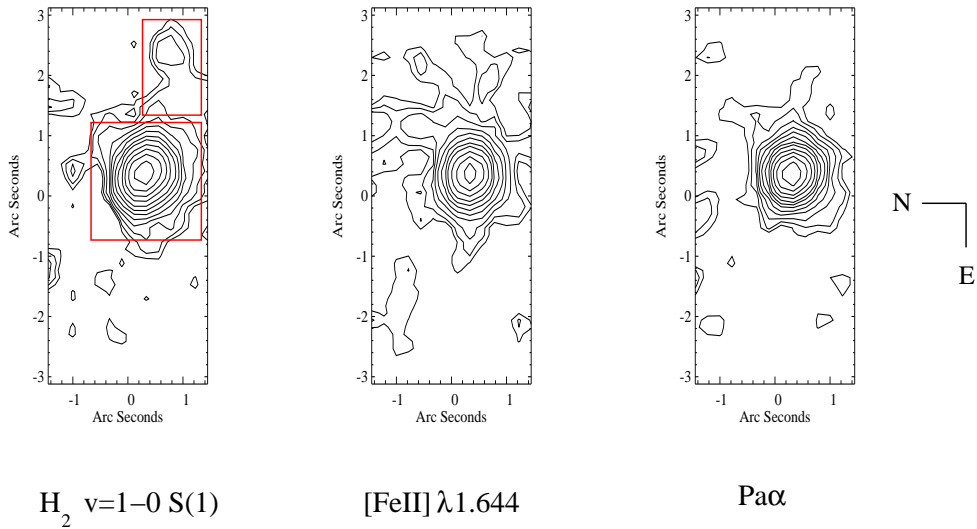}
\caption{\normalsize Contour plots of the \H2~v=1-0~S(1), [FeII]$\lambda1.644$ and Pa$\alpha$ emission derived 
from continuum-subtracted narrow-band cuts on the data cube obtained with the HK grism. The narrow-band images were 
smoothed with 0.48\arcsec~square top-hat filters prior to contouring. The lowest contour levels are set at $2\sigma$ of the background in the lower part of each image, and have values of $3.1 \times 10^{-20}$, $5.2 \times 10^{-20}$ and 
$1.1 \times 10^{-19}$~W~m$^{-2}$, for \H2~v=1-0~S(1), [FeII]$\lambda1.644$ and Pa$\alpha$, respectively. Successive 
contour levels increase by $\sqrt{2}$. The boxes used for extraction of spectra for the nucleus and a filamentary
structure are indicated on the \H2~v=1-0~S(1) map. The orientation on all three panels is as indicated.}
\label{fig:emlines}
\end{figure*}

We began by deriving from the HK data cube emission line maps of \H2~v=1-0~S(1), 
[FeII]$\lambda$1.644 and Pa$\alpha$, as shown in Fig.~\ref{fig:emlines}. They were formed by taking slices through 
the data cube at the wavelengths of interest and subtracting the underlying continuum by using line-free regions either side of the line and interpolating the continuum linearly across the line. The detailed spectral decomposition of the continuum by Krabbe et al.~(2000) shows that there are no stellar absorption features in the vicinity of these emission lines, and that our continuum subtraction procedure is therefore reliable. The maps show that
the emission in all three lines is heavily dominated by the nucleus, and although there is noticeably more 
extended emission to the west of it than to the east, the `filament' of \H2~is the only feature that can be
readily isolated. Its appearance in the narrow-band \H2~image in Donahue et al.~(2000) attests to its reality. 
The \H2~v=1-0~S(1) luminosities of this filament and the $2 \times 2$\arcsec~nuclear region are $1.8 \times 10^{39}$
and $3.0 \times 10^{40}$\ergps, respectively.   A comparison of our measurement of the \H2~v=1-0~S(1) flux in the
$2 \times 2$\arcsec~nuclear region ($43.0 \pm 0.9 \times 10^{-15}$\ergpcmsqps) with that measured in the central 3\arcsec~by Krabbe et al.~(2000) ($41.1 \pm 2.5 \times 10^{-15}$\ergpcmsqps), demonstrates that there has been no variation in the 8.7 years between the two observations. 

\begin{figure}
\includegraphics[width=0.48\textwidth,angle=0]{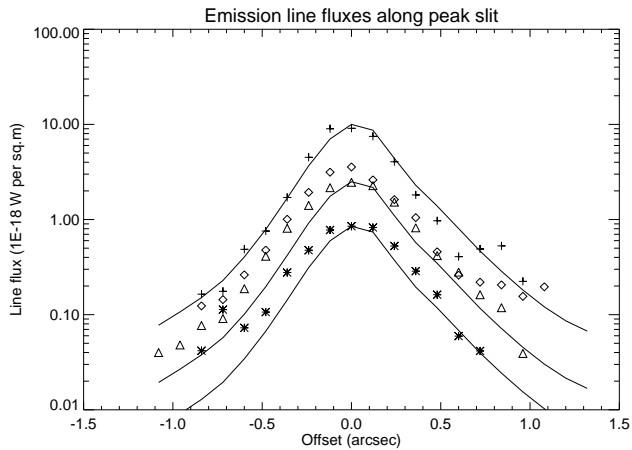}
\caption{\normalsize Variation in emission line flux along the IFU slit passing through the nucleus, for
Pa$\alpha$ (crosses), [FeII]$\lambda$1.644 (diamonds), \H2~v=1-0~S(1) (triangles) and S(2) (asterisks). The lines
show scaled PSFs for this slit, determined from a standard star observation. Line fluxes were derived from gaussian fits to the profiles in the HK data cube.}
\label{fig:fluxes}
\end{figure}

The fine pixel scale of the UIST IFU, coupled with the excellent seeing during the observations, allows us to 
examine the intensity profiles of the emission within 1\arcsec~of the nucleus and to constrain the spatial 
locations of the emitting regions. Almost 80 and 90 per cent, respectively, of the line and continuum emission in this 
central region falls in slits 9 and 10, the bulk of it in slit 9, which we refer to hereafter as the `peak slit'. In 
Fig.~\ref{fig:fluxes} we show intensity profiles along the peak slit for several emission lines, with the fluxes derived
from single gaussian fits to the HK data. The pointspread function (PSF) for this slit, as derived from a 
reconstructed white-light image of a standard star, is also indicated. It can be seen that Pa$\alpha$ is unresolved,
whereas the \H2~and [FeII] emission are both extended with respect to the PSF. 

\begin{figure}
\includegraphics[width=0.48\textwidth,angle=0]{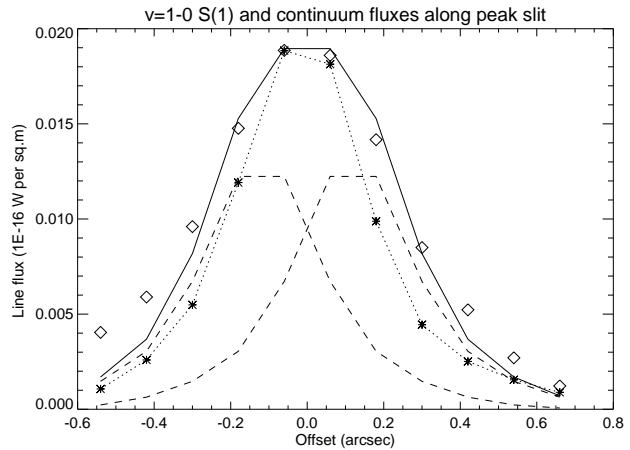}
\caption{\normalsize The diamonds show the \H2~v=1-0 S(1) flux in pixels along the peak IFU row measured relative to the continuum nucleus (negative offsets are to the east). The dashed 
lines show the contributions from two point-sources of emission at $\pm 0.12$\arcsec, and the solid
line their sum. The latter accounts for the bulk of the emission in the central 0.6\arcsec. Line fluxes were derived
from double gaussian fits to the profiles in the short-K data (see section 3.3). The asterisks connected with the dotted line show (with arbitrary normalization) the continuum level below the \H2~v=1-0 S(1) line: it is clearly less extended than the line emission and spatially unresolved.}
\label{fig:fluxmodel}
\end{figure}

A closer look at the \H2~profile is 
shown in Fig.~\ref{fig:fluxmodel}, where the fluxes were derived from double gaussian fits to the profiles from 
the higher spectral resolution, short-K, data cube. They show that the \H2~profile~has a $\rm{FWHM}=0.6$\arcsec, corresponding to a PSF-deconvolved $\rm{FWHM}=0.4$\arcsec. Indeed, Fig.~\ref{fig:fluxmodel} also demonstrates (but does not of itself prove) that the bulk of the \H2~emission out to radii $\simeq 0.4$\arcsec~can be accounted for by a superposition of 2 
point-sources of emission, each located at a radius of approximately 0.15\arcsec~from the continuum peak (i.e. the nucleus), 
which lies between the central two pixels. In contrast, Fig.~\ref{fig:fluxmodel} also demonstrates than the continuum beneath 
the line -- dominated by AGN emission in this central region -- is spatially unresolved. Comparison of the continuum fluxes
in slits 9 and 10 (not shown) implies that the seeing disk of the unresolved point source is offset by $\sim 0.06$\arcsec~from the centre of slit 9, towards slit 10. This reflects the precision with which a source can be acquired within the IFU aperture and hence also limits our determination of the PSF from a single standard star observation. 

To summarise our analysis of the emission line morphologies, we adopt as a working hypothesis a model in which the bulk of the \H2~emission arises in an extended structure located $\simeq 0.15$\arcsec~(50\pc) either side of the nucleus along an east-west axis. Further support for this model comes from an analysis of both the excitation and kinematics of the \H2~emission, which we pursue below.

\subsection{\H2~excitation}
\begin{figure*}
\includegraphics[width=0.44\textwidth,angle=0]{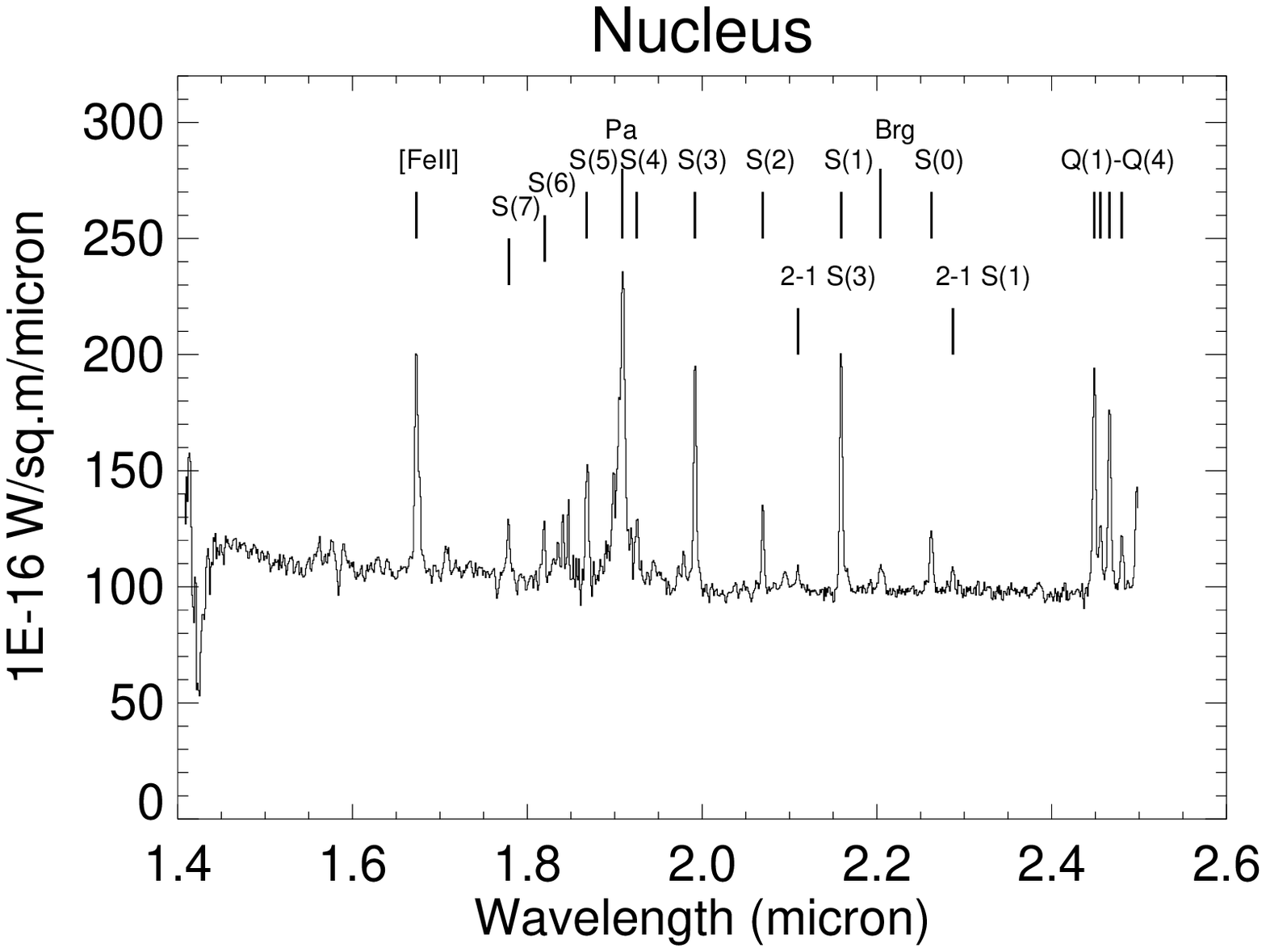}
\includegraphics[width=0.44\textwidth,angle=0]{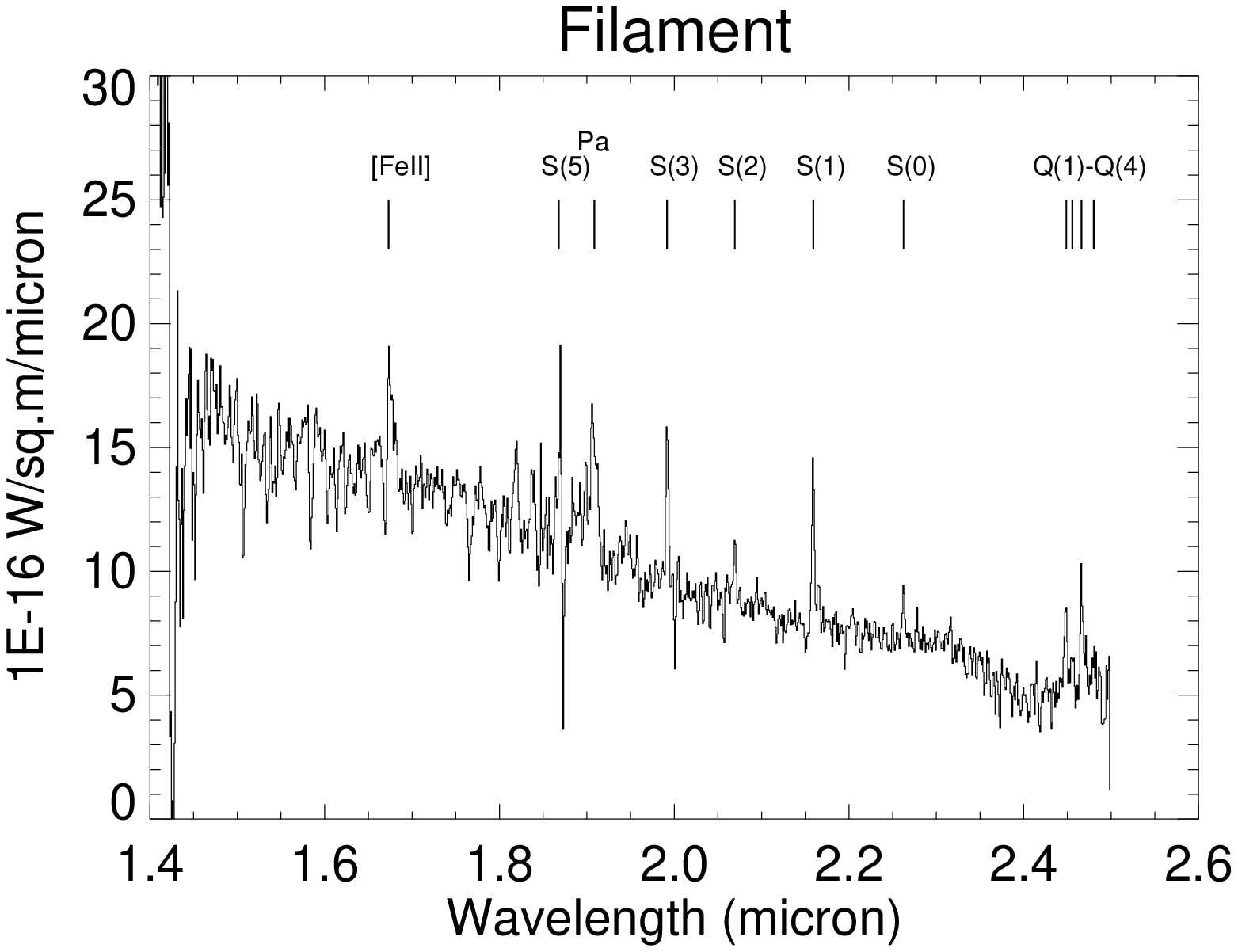}
\includegraphics[width=0.44\textwidth,angle=0]{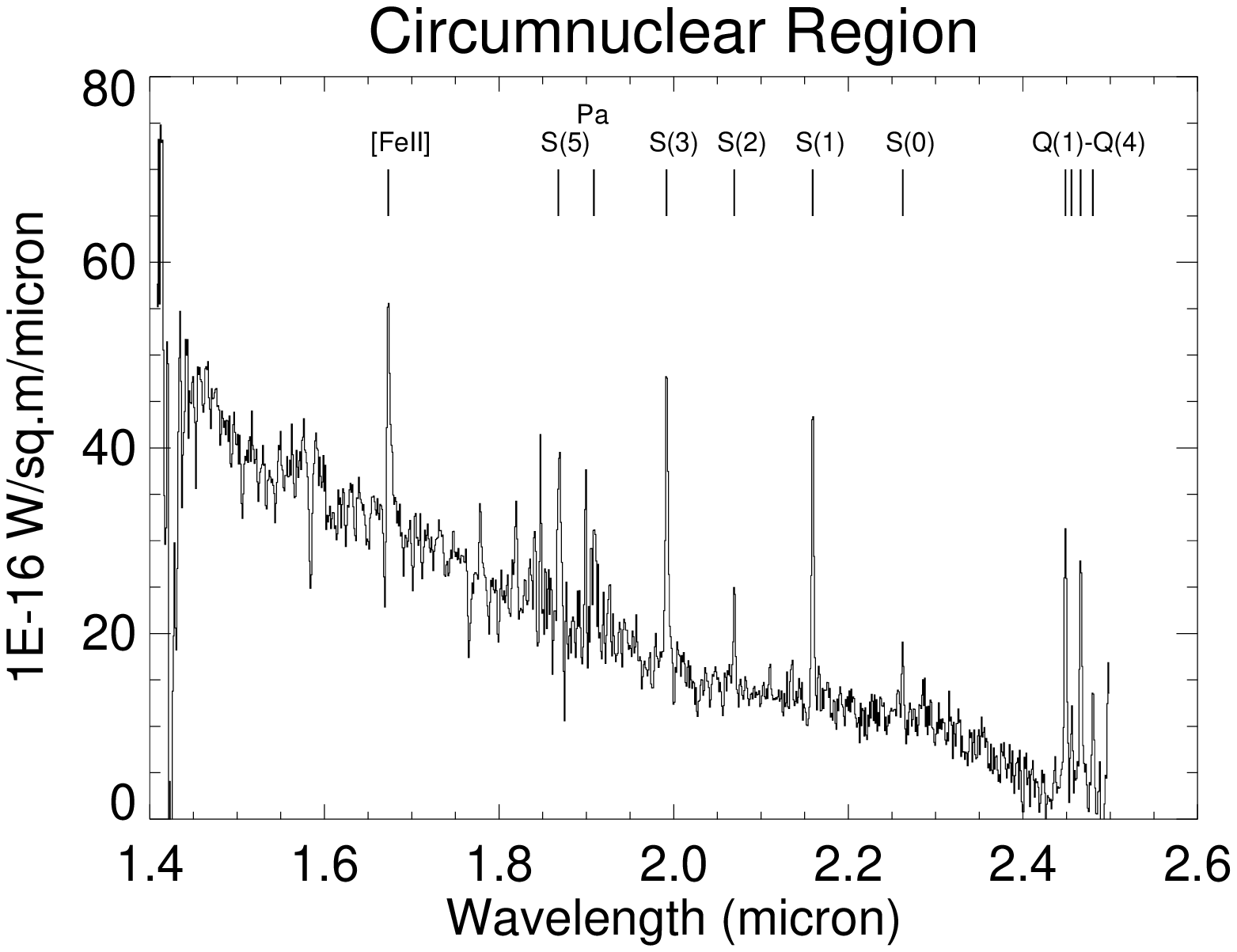}
\caption{HK spectra of the nucleus, the filament and circumnuclear region in NGC 1275. The boxes used for the extraction of the nucleus and filament spectra are shown in Fig.~\ref{fig:emlines}; the spectrum of the circumnuclear region was obtained by subtracting from the $2 \times 2$\arcsec~nuclear box an appropriately scaled version of the spectrum of the central 9 pixels, such that only 
the extended emission in this region remains. Note the broad base to the Pa$\alpha$~line on nucleus, consistent with the Seyfert 1 optical classification of NGC 1275.}
\label{fig:spectra}
\end{figure*}

In Fig.~\ref{fig:spectra} we show HK spectra of the nucleus and the filament referred to in Fig.~\ref{fig:emlines}.
Also shown is the spectrum of the extended emission within the $2 \times 2$\arcsec~nuclear box, formed by subtracting 
a scaled version of the spectrum of the central 9 pixels (scaled using a PSF observation of a standard 
star to take account of light from the point source scattered into other pixels). Although it falls in a region 
of bad atmospheric cancellation, Pa$\alpha$ appears to be very weak in the circumnuclear spectrum, confirming the 
result of Fig.~\ref{fig:fluxes} that it is spatially unresolved. 

To examine the excitation mechanism for the \H2, we begin by comparing the observed \H2~line ratios with the predictions
for thermal (collisional) excitation, which dominates in molecular gas with density $n_{\rm{T}} > 10^{5}$\pcm~(the critical
density of these transitions), heated to a few thousand Kelvin by UV or X-radiation, or in shocks. Under these conditions,
the occupation numbers of the excited ro-vibrational levels of the \H2~molecule will be in thermal equilibrium at a temperature
$T_{\rm{ex}}$ equal to the kinetic temperature of the gas. Therefore, for a given object, the flux $F_{\rm{i}}$ in the
$i$th emission line will satisfy the relation $log(F_{\rm{i}}\lambda_{\rm{i}}/A_{\rm{i}} g_{\rm{i}}) = constant - T_{\rm{i}}/T_{\rm{ex}}$, where $\lambda_{\rm{i}}$ is the wavelength of the line, and $A_{\rm{i}}$, $T_{\rm{i}}$ and $g_{\rm{i}}$ are the
spontaneous emission coefficient, energy (expressed as a temperature) and statistical weight of the upper level of the
transition (the latter assumes an ortho:para \H2~abundance ratio of 3:1, appropriate where collisions dominate).

\begin{figure*}
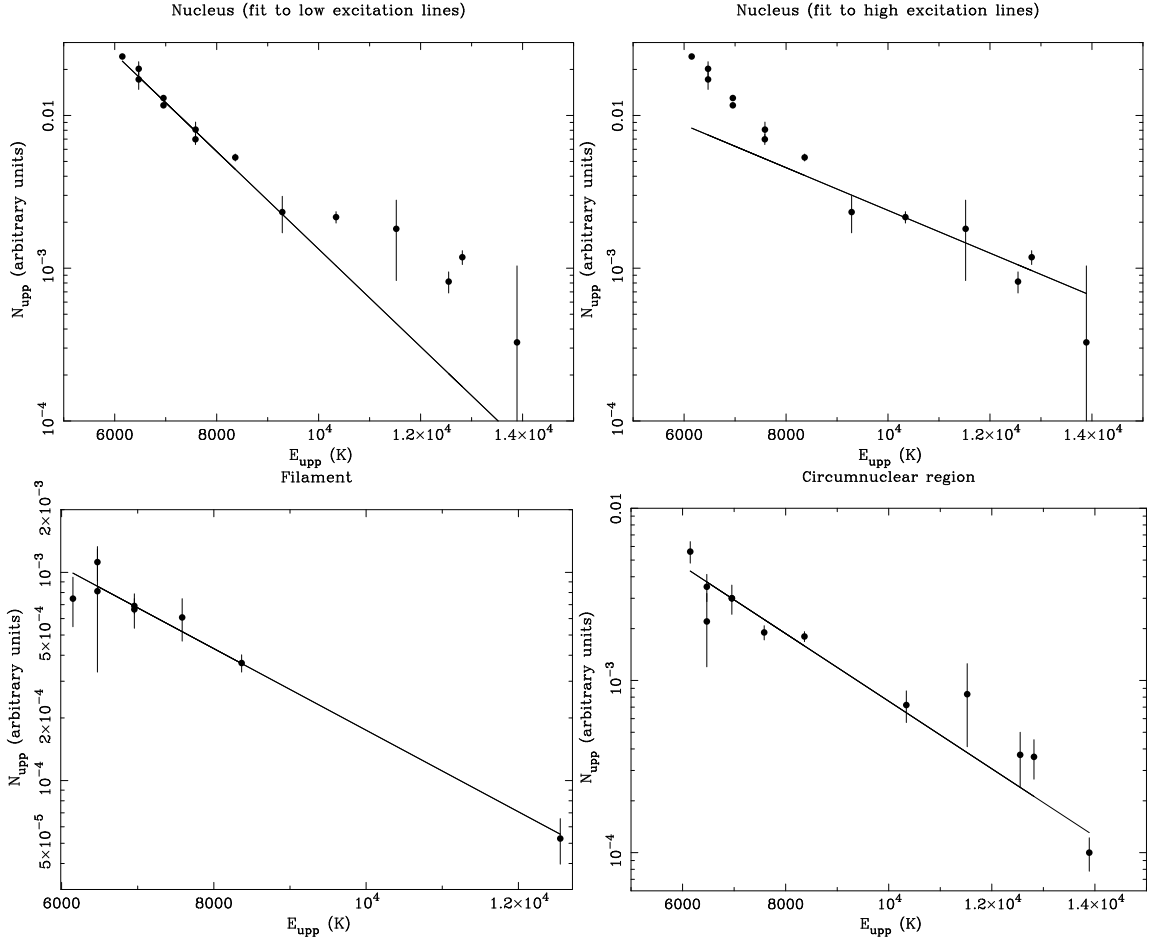

\begin{centering}
\includegraphics[width=0.35\textwidth,angle=270]{nucltelowex_rv.ps}
\includegraphics[width=0.35\textwidth,angle=270]{nucltehiex_rv.ps}
\includegraphics[width=0.35\textwidth,angle=270]{fil_lte.ps}
\includegraphics[width=0.35\textwidth,angle=270]{extendedlte.ps}
\caption{LTE diagnostic plots for the nucleus, the filament and circumnuclear region. Clockwise from top left, the superimposed 
models are for single temperature thermal excitation at temperatures of $T_{\rm{ex}}=1360 \pm 50$\K~(fit just to ($E_{\rm{upper}} 
< 10000$\K~lines), $3100 \pm 500$\K~(fit just to $E_{\rm{upper}} > 10000$\K~lines), $2200 \pm 100$\K, and $2200 \pm 200$\K.}
\label{fig:LTE}
\end{centering}
\end{figure*}

Such plots were constructed for the nucleus, the circumnuclear region and the filament, and are shown in Fig.~\ref{fig:LTE}.
For the nucleus, we see that the low-lying transitions are completely thermalised, with an excitation temperature of $1360 \pm 50$\K, implying a gas density above $10^{5}$\pcm. The excess flux in the higher-excitation transitions may have one of three possible origins: (a) it could be from a much hotter thermalised gas with $T_{\rm{ex}}=3100 \pm 500$\K~(as deduced from a fit to the high-excitation lines only), close to the \H2~dissociation temperature of $\sim 4000$\K~(fitting instead a two-temperature model with the excitation temperature of the hotter component fixed at 3100\K~implies that the cooler component has a temperature of $800 \pm 50$\K) (b) it could be non-thermal fluorescent emission in low density gas; (c) it could be non-thermal 
emission due to excitation of the molecule by secondary electrons deep in the cloud. We confirm the finding of Krabbe et al.~(2000) that on the basis of model (a) the v=2-1~S(3) transition with $T_{\rm{i}}=13890$\K~is underluminous by a factor of $\sim 3$, explained by the fact that the upper level of this transition is de-populated by an accidental resonance with photons around the wavelength of HI Ly$\alpha$ at 1216\AA~(Black \& van Dishoeck~1987). Our excitation temperature for the low excitation lines is slightly lower than that found by Krabbe et al.~(2000), who measured $1480 \pm 250$\K, but our determination is more precise due to our inclusion of several 1-0~Q series lines above 2.4\micron. For the higher excitation lines, Krabbe et al.~determined temperatures between 2600 and 2900\K. For the circumnuclear and filament and circumnuclear spectra, we deduce excitation temperatures of $2200$\K, as expected for X-ray or shock-heating.

Given an excitation temperature, the v=1-0~S(1) fluxes can be converted to masses of hot molecular hydrogen in the following manner. For optically thin emission (which is a valid assumption for these transitions), the H$_{2}$ column density, $N(H_{2})$, is related to the observed intensity of the v=1-0~S(1) emission, $I$, by the expression $N(H_{2}) = 4 \pi I / f A h \nu$, where $A=3.47 \times 10^{-7}$\ps, $h \nu = 9.37 \times 10^{-13}$\erg, and $f$ is the fraction of H$_{2}$ molecules in the $v=1, J=3$ state leading to v=1-0~S(1) emission. Integration over the solid angle on the sky covered by the source implies that the mass of hot H$_{2}$ is given by $M(H_{2}) = 4 \pi L_{1-0 S(1)} m_{H_{2}} / f A h \nu$, where $L_{1-0 S(1)}$ is the v=1-0~S(1) luminosity and $m_{H_{2}}$ is the mass of the H$_{2}$ molecule. Computation of the ro-vibrational H$_{2}$ partition function at various temperatures enables us to deduce the appropriate value of $f$, which varies as follows: 0.00031 (800\K), 0.0049 (1360\K), 0.016 (2200\K), 0.023 (3100\K). Hence for the nucleus we deduce the following H$_{2}$ masses: (i) $3.9 \times 10^{5}$\Msun, if all the v=1-0~S(1) is thermal emission at 1360\K; (ii) $1.0 \times 10^{6}$\Msun~at 800\K~and $4.5 \times 10^{4}$\Msun~at 3100\K, in the aforementioned two-temperature model. For the circumnuclear and filament spectra, the masses of hot H$_{2}$~at 2200\K~are $3.3 \times 10^{4}$\Msun~and $7.3 \times 10^{3}$\Msun, respectively.

With reference to the work of Maloney, Hollenbach \& Tielens~(1996) we now assess the energetic requirements for X-ray heating of the molecular gas. As discussed therein, the dominant parameter controlling the physical conditions is $H_{\rm{X}}/n$, the ratio of
the X-ray energy deposition rate per particle, $H_{\rm{X}}$, to the gas density, n. The former is the integral over energy of 
$F(E)\sigma_{\rm{pe}}$, where $F(E)$ is the local photon energy flux per unit energy interval and $\sigma_{\rm{pe}}$ the photoelectric cross-section per hydrogen atom. For an X-ray source with a power-law spectrum of photon index $\Gamma = 2$ and 1--100\keV~luminosity $L_{\rm{X}}=10^{44}L_{\rm{44}}$\ergps, located $50r_{\rm{50}}$\pc~from the gas cloud:

\begin{equation}
H_{\rm{X}} \sim 3 \times 10^{-21} L_{\rm{44}} r_{\rm{50}}^{-2} N_{\rm{22}}^{-1} \ergps
\end{equation}

where $N_{\rm{H}}=10^{22}N_{\rm{22}}$\psqcm~is the equivalent neutral hydrogen column density attenuating the X-ray flux; a column of at least $10^{21}$\psqcm~is required in order to exclude the normal UV photon-dominated region at the cloud surface. Applied to NGC~1275, we find that:

\begin{equation}
log H_{\rm{X}}/n \simeq -25.3 + log[L_{\rm{X}}/L_{\rm{Einstein}}] - log[r_{\rm{50}}^{2} n_{\rm{5}} N_{\rm{22}}].
\end{equation}

Here $L_{\rm{X}}$ is the X-ray luminosity of the source as seen by the cloud and $L_{\rm{Einstein}} = 1.6 \times 10^{44}$\ergps~is the X-ray luminosity of the source extrapolated (in energy) from the measurement by the Einstein satellite in 1979~(Branduardi-Raymont et al.~1981), which also suggests an obscuring column $N_{\rm{22}} \sim 1$; $n_{\rm{5}}$ is the gas density in units of $10^{5}$\pcm, which as mentioned above must be unity or higher for thermal excitation. The modelling of Maloney et al.~(1996) demonstrates that the bulk of the thermally excited \H2~emission is produced over a fairly narrow range around $log H_{\rm{X}}/n = -26$, with the $\H2$ emissivity dropping by at least a factor of 3 when $log H_{\rm{X}}/n$ is more than $\pm 0.3$~dex from the emissivity peak. Equation (2) therefore demonstrates that X-ray heating is an energetically feasible excitation mechanism for the observed thermal \H2~emission. It is, however, difficult to make a more precise statement because of the uncertainy over the appropriate value of $L_{\rm{X}}$, since the light travel time from the nucleus to the clouds is $\sim 160$\yr, but observations of the source over just the last 30 years show that $L_{\rm{X}}$ has been far from constant. From the early-1970s to the mid-1990s the X-ray luminosity of the central source has dropped steadily by a factor of 20 (see e.g. Levinson et al.~1995 for a summary of the observational evidence). It is, however, clear that unless the AGN were much weaker in the past or the gas significantly denser than $n_{\rm{5}} \sim 1$, the hot \H2~could not exist much within the observed 50\pc~radius. Similarly, equation (2) also dictates that, all other parameters being equal, thermally excited \H2~emission is produced in fairly narrow range of radii, within $\pm 0.15$~dex of the peak radius, $\sim 35-70$\pc~in this case. 

In concluding this subsection, we point out that the results of Maloney et al.~(1996) for X-ray heating assume an equilibrium state, which is established on a time-scale of $t_{\rm{eq}} \sim 750 L_{\rm{44}}^{-1} r_{\rm{50}}^{2}$\yr~(set by the ionization rate of hydrogen, which has the slowest chemistry to equilibriate). Therefore, if the central X-ray source is varying significantly on timescales of $t_{\rm{eq}}$ or less -- and observations over the last 3 decades suggest that it is -- consideration of non-equilibrium chemistry may be necessary.

\subsection{\H2~kinematics: evidence for a rotationally-supported structure}
We turn our attention now to the kinematics of the molecular gas in the vicinity of the nucleus, as derived from the
short-K data. The most striking feature, visible in the raw data, is the strong velocity shear in the peak position of the 
\H2~v=1-0~S(1) line moving across the nucleus. To demonstrate this, we show in Fig.~\ref{fig:S1lines} line profiles on a 
pixel-by-pixel basis along the peak slit of the IFU, passing through the nucleus. The kinematics were quantified by fitting
double gaussian profiles to the lines, with the widths of the two components constrained to be the same. The fitting of such
profiles does not imply that gas of both velocities is present at each physical location in the galaxy; rather, the weaker of
the two components is simply scattered light from gas on the other side of the nucleus, in line with the model of 
Fig.~\ref{fig:fluxmodel} in which the bulk of the \H2~is concentrated in two approximately point-like components 
0.15\arcsec~either side of the nucleus. Further support for the latter model is provided by Fig.~\ref{fig:vel}, which shows 
the fluxes of the fitted red and blue components in the peak slit (slit 9) and also in slit 10. We also show rotation curves 
for the central four slits (i.e. slits 8--11); for slits 9 and 10, the velocity being that of the brighter of the two 
gaussians at each location. For slits 8 and 11, the lines are fainter, so only a single gaussian can be fitted to these lines.
The appearance of these rotation curves, with the sharp decrease in velocity amplitude in the slits either side of the nucleus suggests that the molecular gas may be in a disk-like structure, rotating about an axis oriented approximately north-south, 
coincident with the radio-jets.

What is the nature of this rotating \H2~structure? It is unlikely to be part of an optically-thick torus of the type commonly 
postulated in the centres of Seyfert galaxies (e.g. Antonucci et al.~1993), although there is evidence for such a 
parsec-scale torus in NGC 1275. Firstly, the presence of a hot dust continuum in the K-band implies that some dust is heated 
to its sublimation temperature and thus lies as close as 0.05\pc~to the nucleus (Krabbe et al.~2000). Secondly, gas in a 
parsec-scale torus perpendicular to the radio jet axis is a plausible candidate for producing the observed free-free 
absorption towards the milliarcsec-scale northern radio counter-jet in NGC 1275 (Levinson, Laor \& Vermeulen~1995); with reference to the latter's models, constraints since obtained on the black hole mass in NGC 1275 (see subsection 3.5) 
suggest that the torus is comprised of individual clumps of dense gas ($n \gg 10^{4}$\pcm) and is matter-bounded in the 
radial direction. At larger radii, NGC 1275 has a complex dust distribution extending out to 17~kpc, in contrast to the 
majority of FRI radio sources in which the dust is settled in disks with radii $<2.5$\kpc~(de Koff et al.~2000)

A more likely possibility is that the hot \H2~is part of a disk, from which we only see emission 
from a ring covering a narrow range of radii around 50\pc, due to the nature of the X-ray heating by the central source 
(see section 3.2). Would such a disk be stable? We consider first a gas disk, in vertical hydrostatic equilibrium in the 
gravitational field of a central black hole of mass $M$. To ensure that the disk is stable against self-gravity we 
require $\Sigma < \frac{M h}{4 r^{3}}$, where $\Sigma$ is its surface mass density and $h$ its exponential scale-height 
(this is essentially the Toomre~(1964) stability criterion). There are two approaches to evaluating this criterion, leading 
ultimately to the same conclusion: (i) integrating vertically over the disk density profile, $n = n_{\rm{0}} 
exp(-z^{2}/h^{2})$, we find $\Sigma = 2 \sqrt{\pi} n_{\rm{0}} h m_{\rm{p}}$. Hence the stability criterion becomes 
$r^{3} < \frac{M}{8 n_{\rm{0}} m_{\rm{p}} \sqrt{\pi}}$. The mid-plane density, $n_{\rm{0}}$ must be at least $10^{5}$\pcm~for
thermal \H2~excitation, and taking $M=3.4 \times 10^{8}$\Msun~as the black hole mass~(see section 3.5), this translates into 
$r < 21$\pc~i.e. less than the observed radius of the \H2~distribution. (ii) Assuming that the mass of hot nuclear \H2~deduced in section 3.2 ($3.9 \times 10^{5}$\Msun) is spread over an annulus from 35--70\pc~we deduce $\Sigma = 0.07$~Kg~m$^{-2}$, so re-casting the stability criterion with $r=50$\pc~we require $h/r > 10^{-3}$. However, from above $\Sigma = 2 \sqrt{\pi} n_{\rm{0}} h m_{\rm{p}}$, so requiring $n_{\rm{0}} > 10^{5}$\pcm~we deduce that $h/r < 8 \times 10^{-5}$, once again inconsistent 
with the stability criterion. We therefore conclude that a smooth, stable, gaseous disk of the observed radius and mass density cannot exist. The propensity for 100\pc-scale nuclear gas disks to be subject to gravitational instabilities was highlighted recently by Tan \& Blackman~(2004) in the context of accretion onto the nuclei of giant elliptical galaxies. The disk can, however, be stabilised if the gas is located in dense clouds, as opposed to being smoothly distributed. In this case, the Toomre stability criterion can be cast as $\frac{\sigma \Omega}{G \Sigma} > 1.68$, where $\sigma$ is the 1-d velocity dispersion of the
clouds and $\Omega = \sqrt{G M /r^{3}}$ the orbital angular velocity. Assuming $\Sigma  = 0.07$~Kg~m$^{-2}$ as before, we hence require $\sigma > 70$\m\ps~for stability, which is not at all restrictive.

From the short-K data we also investigated the kinematics of the \H2~filament. With respect to the assumed velocity 
zero-point in Fig.~\ref{fig:vel}, the \H2~velocity of this gas is $-10 \pm 14$\kmps, with a de-convolved linewidth of 190\kmps~FWHM. Hence, this material is unlikely to be part of the putative disk. The wavelength range of the short-K data does not 
contain [FeII] or Pa$\alpha$ so we are unable to study the kinematics of these lines in comparable detail.

In this subsection we have demonstrated that the \H2~is plausibly in a disk-like structure, composed of individual clumps of 
dense gas since a smooth gas disk of the observed dimensions would be gravitationally unstable. Because the points in the 
rotation curves are not mutually independent -- the pixel separation along each slit is 0.12\arcsec~whilst the seeing is $\simeq 0.4$\arcsec~--  we have not attempted to fit a full two-dimensional disk model to the data, since this would require a much better knowledge of the PSF than we have. Adaptive optics observations are needed to conclusively test the disk model in two dimensions. Nevertheless, in the next section we assume that the disk model is valid and make an estimate of the black hole mass.

\subsection{Black hole mass estimation}
The computation of dynamical black hole mass estimates for galactic nuclei has reached a considerable level of complexity, 
requiring in general observations of the full two dimensional velocity and velocity dispersion fields, knowledge of the distributed 
stellar mass and dynamical modelling. The resulting model of the gas kinematics must then be projected on to the plane of the sky, 
convolved with the instrumental response and fitted to the observational data. Here we circumvent all of this complexity and
compute a simple estimate of the mass enclosed within the observed radius of the molecular gas. Thereafter, we use an archival
HST NICMOS image of NGC 1275 to estimate the stellar mass contribution in this region. The resulting black hole mass is then found
to compare favourably with other non-dynamical estimates.

From the rotation curve for the nuclear slit (slit 9) shown in Fig.~\ref{fig:vel}, we see that the \H2~v=1-0~S(1) velocity
jumps by approximately $240$\kmps~across the nucleus. If the perpendicular to the disk is inclined at an angle $i$ to the plane of the
sky the velocity difference increases to $\Delta V = 240/ \rm{sin}~${\it i}~\kmps. Our best estimate of $i$ derives from VLBI radio observations of the parsec-scale jets, which suggest that the radio jets are inclined to the line of sight at an angle of 30--55 degrees, with the southern jet approaching us (Walker, Romney \& Benson~1994). On the assumption that the molecular gas rotation axis coincides with the radio jet axis we take $i=45$~degrees. Within the inferred molecular gas radius of $r=50$\pc, the dynamical mass is given by:

\begin{equation}
M(total) = \frac{v^{2} r}{G} 
\end{equation}

with $v=\Delta V/2 = 120/\rm{sin}$~{\it i}~\kmps. The result is $M(total)=3.4 \times 10^{8}$\Msun. Taking account of the aforementioned uncertainty in the inclination of the system introduces a systematic uncertainty of $\pm 0.18$~dex on this mass.

\begin{figure}
\includegraphics[width=0.48\textwidth,angle=0]{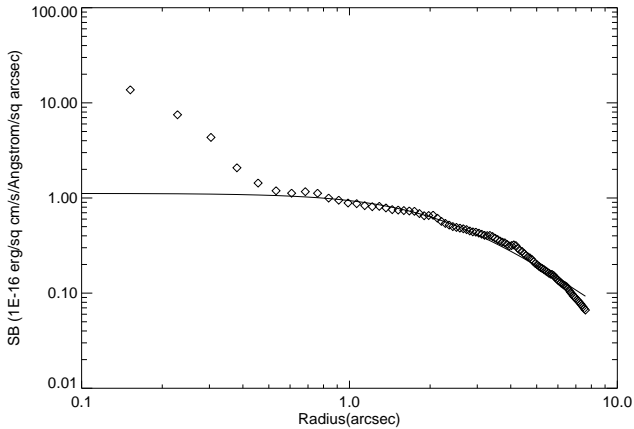}
\caption{\normalsize The surface brightness of the inner regions of 
NGC 1275 as function of the radius of the semi-major axis of the elliptical isophotes, derived from an HST NICMOS observation in 
the F160W filter (corresponding closely to the H-band). The line shows the best fit modified Hubble profile at radii $>0.5$\arcsec, with the upturn in the data below this radius being caused by non-stellar emission associated with the AGN.}
\label{fig:nicmos}
\end{figure}

For the next stage of the calculation we used an image of NGC 1275 to infer the stellar mass distribution. The data in question are 
a 640 second exposure HST NICMOS image taken in the F160W filter (roughly corresponding to the H-band) on 03/16/98 (dataset number 
N3ZB1R010). The standard pipeline-processed dataset was obtained from the archive. Using the IRAF STSDAS task $ellipse$, elliptical
isophotes were fitted and the resulting surface brightness profile is shown in Fig.~\ref{fig:nicmos}. Since the eccentricity of the
isophotes is typically small ($e \sim 0.2$) we chose to fit the profile with a spherically symmetric modified Hubble Law:

\begin{equation}
I(R) = \frac{I_{0}}{1 + (R/R_{0})^2}
\end{equation}

which deprojects analytically to the following spatial luminosity density:

\begin{equation}
j(r) = \frac{j_{0}}{[1 + (r/R_{0})^2]^{3/2}}.
\end{equation}

Fitting to the data at radii $> 0.5$\arcsec~(beyond the region dominated by the AGN power-law and re-radiated dust emission; see Krabbe et al.~2000 for a spectral decomposition of the continuum) we deduce $R_{0}=2.3$\arcsec~and $I_{0}=1.12 \times 10^{-16}$erg~cm$^{-2}$~s$^{-1}$~\AA$^{-1}$~arcsec$^{-2}$. Although there are some systematic deviations from the model at $R>3$\arcsec~it is certainly good to within a factor of 2. When deprojected, the implied central stellar mass density is
6.3(M/L)$_{H}$\MsunMpc, where $(M/L)_{H}$ is the mass-to-light ratio of the stellar population in solar units. Values of $(M/L)_{H} \sim 1$ are expected for stellar populations with age $\sim 10$~Gyr for a wide range of power-law and exponential stellar initial mass functions and metallicities (Salasnich et al.~2000); also, for a given population, $(M/L)_{H}$ will vary in the approximate range 0.05--3 as the population evolves, and will in general increase with age. Assuming, therefore, that $(M/L)_{H}=1$, the implied stellar mass within the 50\pc~radius of the molecular gas distribution is $3.7 \times 10^{6}$\Msun, which is $\simeq 1$~per cent of the above figure for $M(total)$. Hence our estimate of the black hole mass is $M(BH)=3.4 \times 10^{8}$\Msun, $\pm 0.18$~dex.

It is interesting to compare this figure with other estimates of the black hole mass in NGC 1275. 
Bettoni et al.~(2003) used the observed central stellar velocity dispersion ($\sigma = 250$\kmps) in conjunction with the
established $M(BH)-\sigma$ correlations (Gebhardt et al.~2000; Ferrarese \& Merritt~2000) to deduce $M(BH)=4.1 \times 
10^{8}$\Msun, $\pm 0.4$~dex. Using the $M(BH)-L_{\rm{Bulge}}$ relation (Magorrian et al.~1998), in which the scatter is arguably somewhat higher, they calculate $M(BH)=1.4 \times 10^{9}$\Msun, $\pm 0.4$~dex. For comparison, dynamical estimates of the black hole mass are available for two other nearby central cluster galaxies, Cygnus A ($M(BH)=2.5 \pm 0.7 \times 10^{9}$\Msun; Tadhunter et al.~2003) and M87 ($M(BH)=3.2 \pm 0.9 \times 10^{9}$\Msun; Marconi et al.~1997). NGC~1275 and M87 are both FRI radio sources with virtually the same modest radio power ($log P(178 MHz) =$ 24.6 W~m$^{-2}$~sr$^{-1}$), but the large difference in their black hole masses is consistent with the scatter at low power in the large sample of $z \simeq 0.5$ radio galaxies in McLure et al.~(2004). Cygnus A, in contrast, is a powerful FRII radio source ($log P(178 MHz) =$ 27.7 W~m$^{-2}$~sr$^{-1}$) with a high-excitation
nuclear spectrum, and falls at the high end of the $M(BH)$--radio power correlation for such objects in the McLure et al.~sample. We conclude that our dynamical estimate of the black hole mass in NGC~1275 is consistent with the most reliable existing estimate and that this value is itself in line with expectations based on its radio properties.

\subsection{Relationship between the hot \H2~and the large-scale distribution of CO emission}
Interferometry of the CO J=1-0 line by Inoue et al.~(1996) suggests that some $3 \times 10^{10}$\Msun~of cool \H2~exists 
within the central arcmin of NGC 1275, in the form of a plume extending from the nucleus to 10\kpc~west of it. 
Within the vicinity of the nucleus, $6 \times 10^{9}$\Msun~of this material is confined within a ring-like structure which 
appears on the sky as two peaks: peak 1, 1.2\kpc~west of the nucleus, and peak 2 the same distance to the south east. The gas
appears to be rotating about the nucleus with line-of-sight velocities $\pm 150$\kmps, in the same sense as the \H2~we have discovered on much smaller scales, suggesting a possible dynamical connection between the two gas systems. Inoue
et al.~(1996) suggested that the CO ring could be due to gas trapped at the inner Lindblad resonance (ILR) of the galaxy potential; 
from poorer spatial and spectral resolution near-infrared observations they suggested that the central peak of hot \H2~could be due to shock emission in a turbulent distribution of molecular clumps funnelled inwards from the CO ring. In 
contrast, we find that the small-scale hot \H2~is in a rotationally supported structure and that the nuclear radiation field is the principle source of excitation. 

To further investigate the connection between the CO ring and the hot \H2~disk, it would be of interest to search for \H2~emission coincident with the former (which falls outside out UIST IFU field of view). There is no \H2~emission at this position in the narrow-band HST NICMOS \H2~image of Donahue et al.~(2000), suggesting that any associated emission is fainter than the \H2~`filament' (which they do detect; see our Fig.~\ref{fig:emlines}). This filament may represent (possibly shocked) material 
being transported from the CO ring to the \H2~disk: it lies at the systematic velocity (see section 3.3) and has a thermal 
\H2~excitation characteristic of shock emission (see section 3.2). 

\begin{figure*}
\includegraphics[width=0.44\textwidth,angle=0]{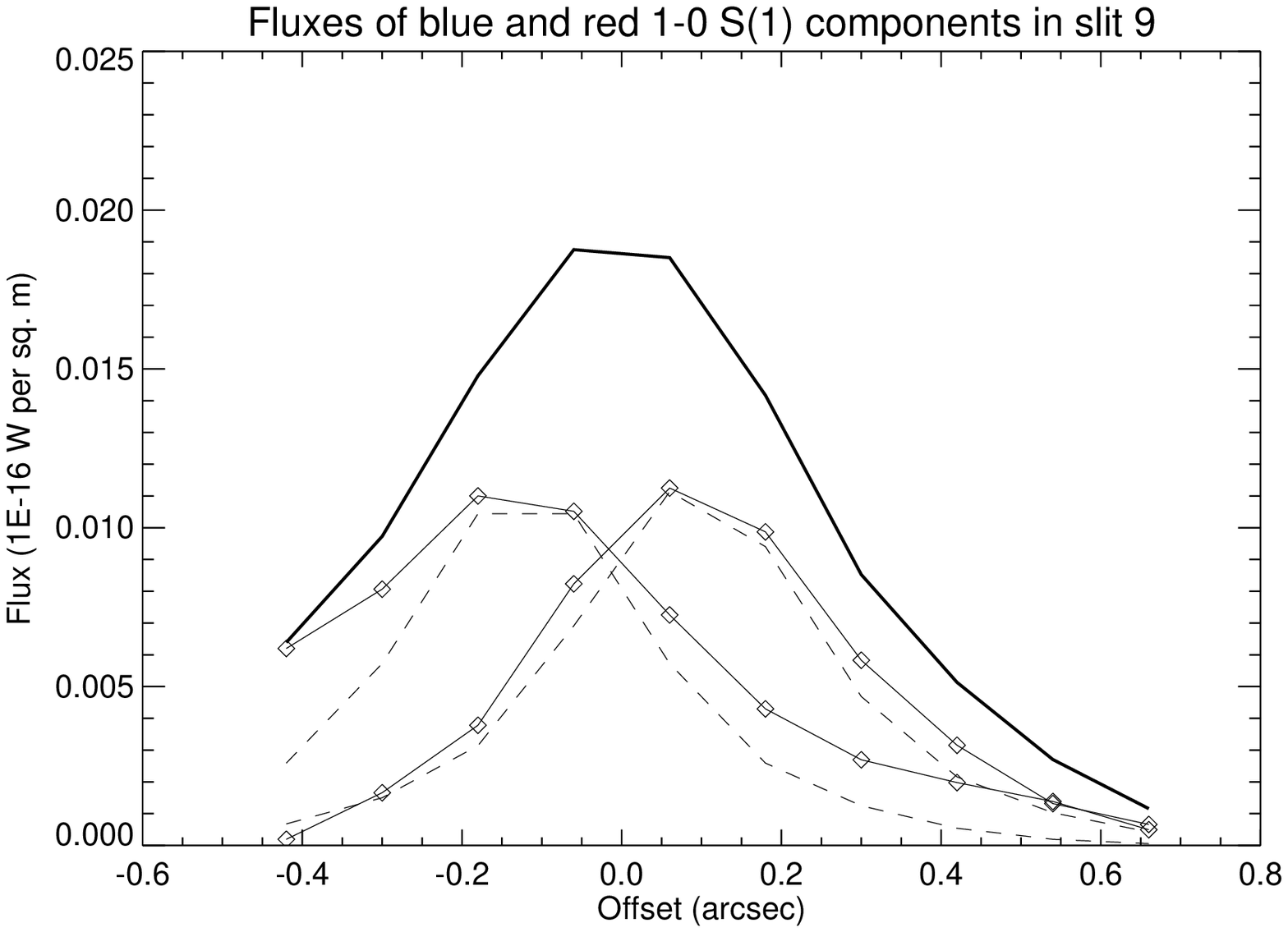}
\includegraphics[width=0.44\textwidth,angle=0]{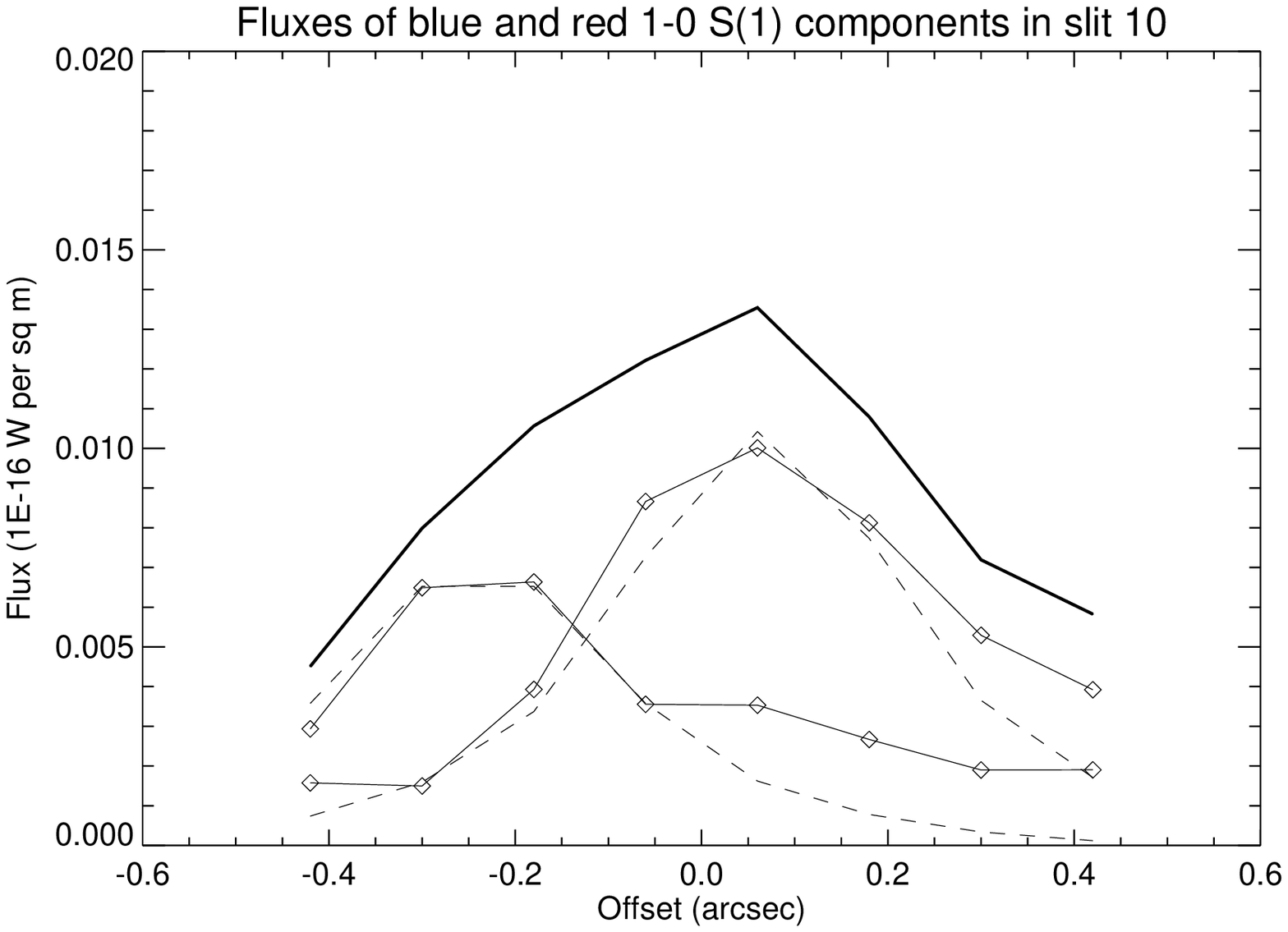}
\includegraphics[width=0.44\textwidth,angle=0]{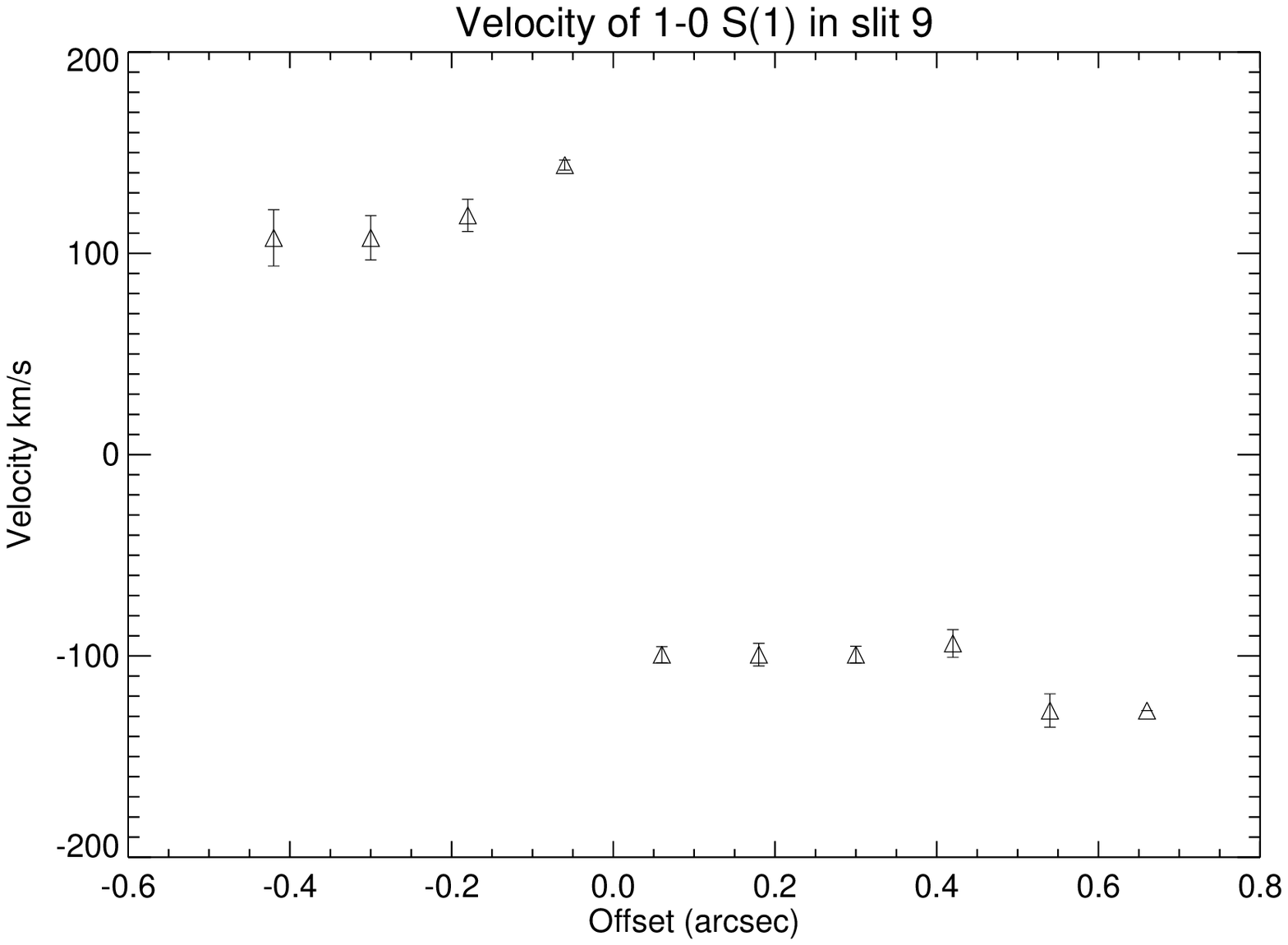}
\includegraphics[width=0.44\textwidth,angle=0]{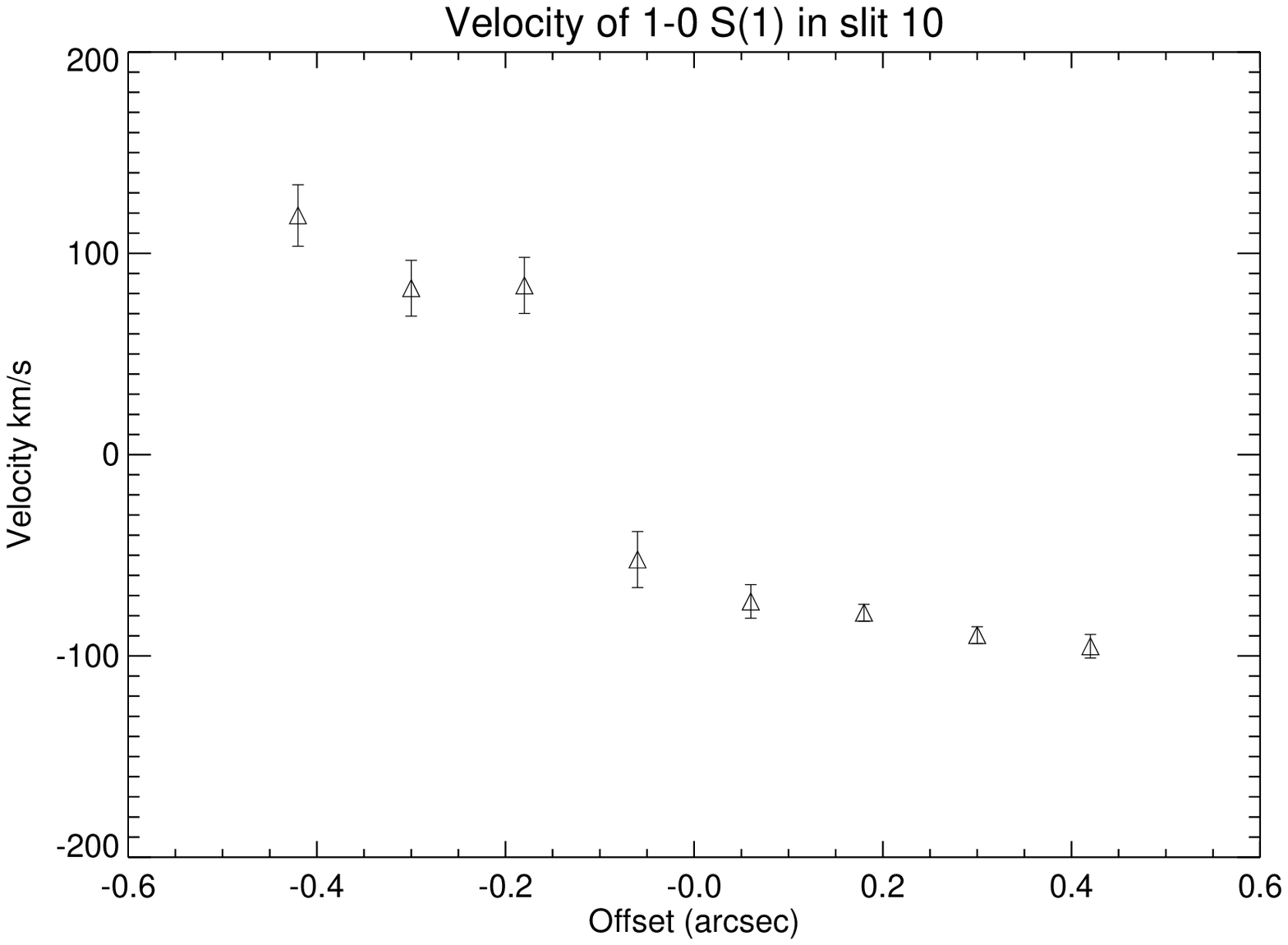}
\includegraphics[width=0.44\textwidth,angle=0]{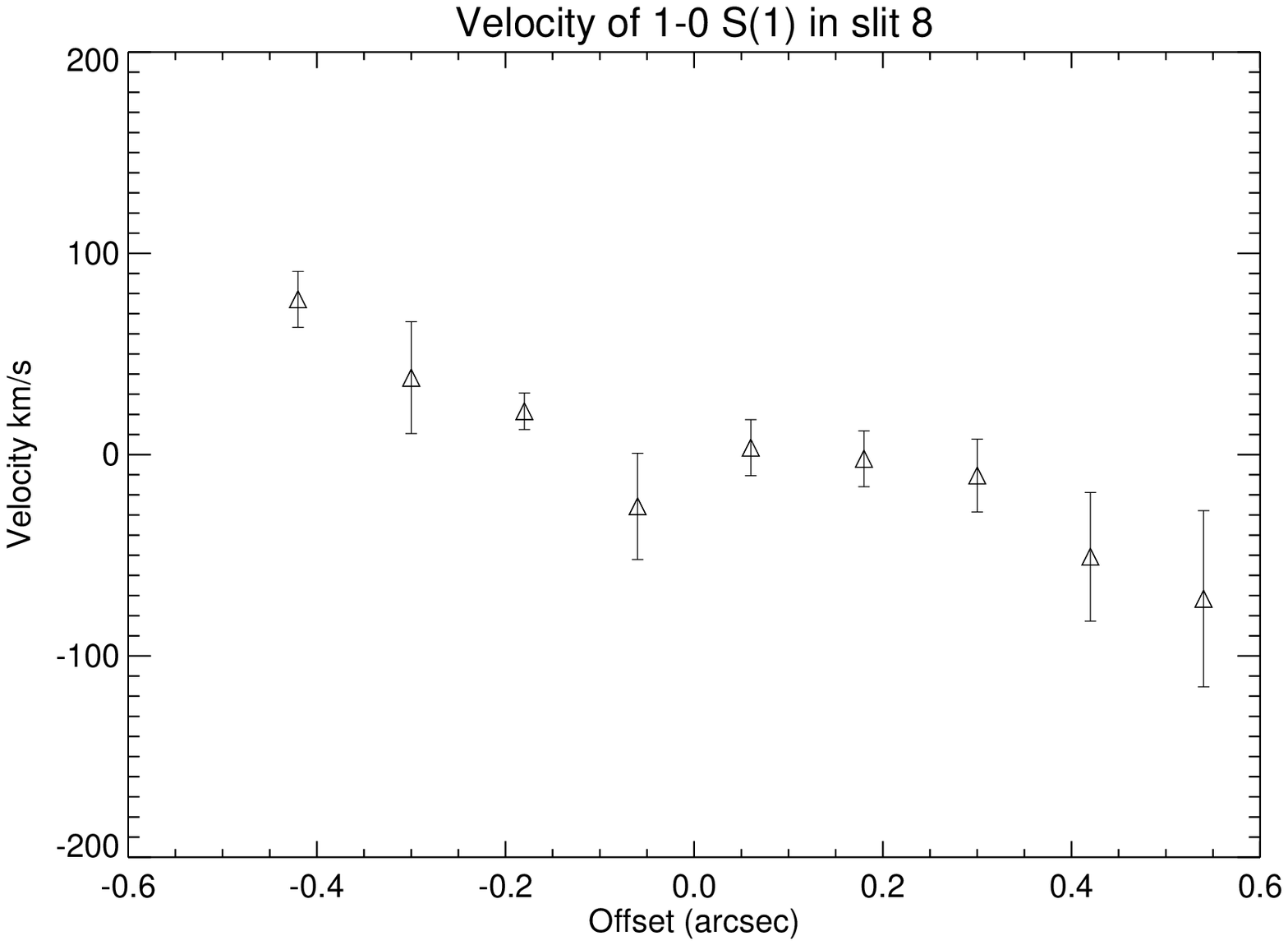}
\includegraphics[width=0.44\textwidth,angle=0]{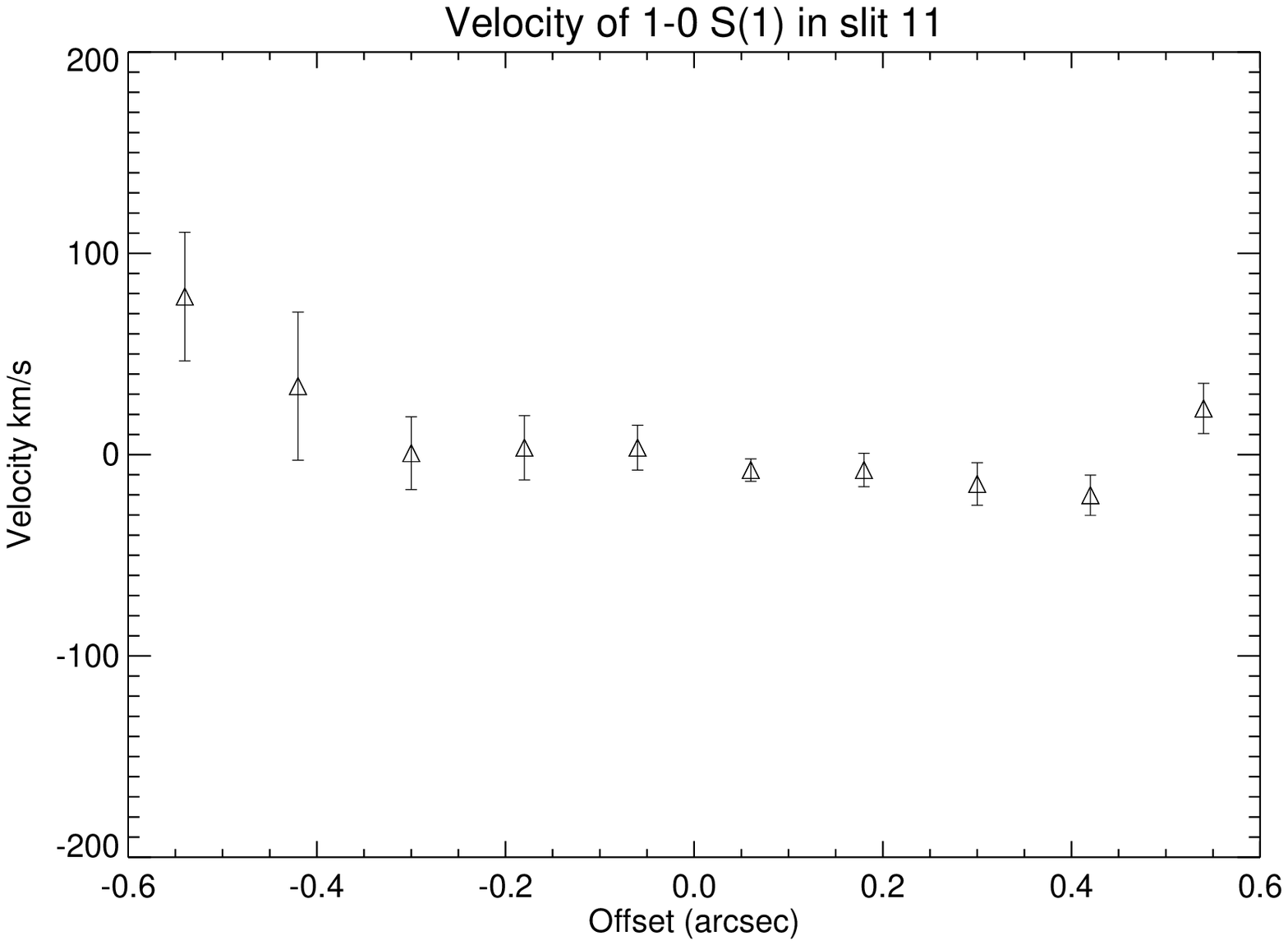}

\caption{Upper left: the diamonds show the fluxes of the red and blue gaussian components fitted to
the \H2~v=1-0~S(1) lines in slit 9 (see Fig.~\ref{fig:S1lines}), and the thick solid line their sum. The dashed lines are PSFs. Upper right: as at upper left but for slit 10 (in both these slits, the blue component is the one which dominates at positive offsets) . Centre left: the velocity of the brighter of the two \H2~v=1-0~S(1) gaussian line components as a function of position for slit 9. Centre right: as at centre left but for slit 10. Lower panels: rotation curves for slits 8 and 11, as derived from single gaussian fits to \H2~v=1-0~S(1). The abscissa on all plots is the offset along the slit relative to pixel position 29.5 (see Fig.~\ref{fig:S1lines}), with negative offsets corresponding to points east of the nucleus. All velocity error bars are 1$\sigma$.}
\label{fig:vel}
\end{figure*}

We also note that the absolute velocity of CO ``peak 1'' 1.2\kpc~west of the nucleus is virtually identical to that of the hot \H2~at a radius of 50\pc~on the same side of the nucleus: on the slit 9 rotation curve of Fig.~\ref{fig:vel}, CO ``peak 1'' would lie at an offset of $+3.5$\arcsec~with velocity $-125$\kmps. In the most naive of dynamical models this would imply that $M(<r)/r$ is the same at radii of 50\pc~and 1.2\kpc. Integrating equation (5), the enclosed stellar masses within these radii are $3.7 \times 10^{6}(M/L)_{H}$ and $1.5 \times 10^{10}(M/L)_{H}$, respectively. Equality in $M(<r)/r$ can be achieved by adding a central black hole of mass $M(BH) = 6.5 \times 10^{8}(M/L)_{H}$\Msun. This shows that the black hole mass deduced from consideration of the hot \H2~dynamics alone is not inconsistent with the assumption that $(M/L)_{H}$ is close to unity.

\begin{figure*}
\includegraphics[width=0.88\textwidth,angle=0]{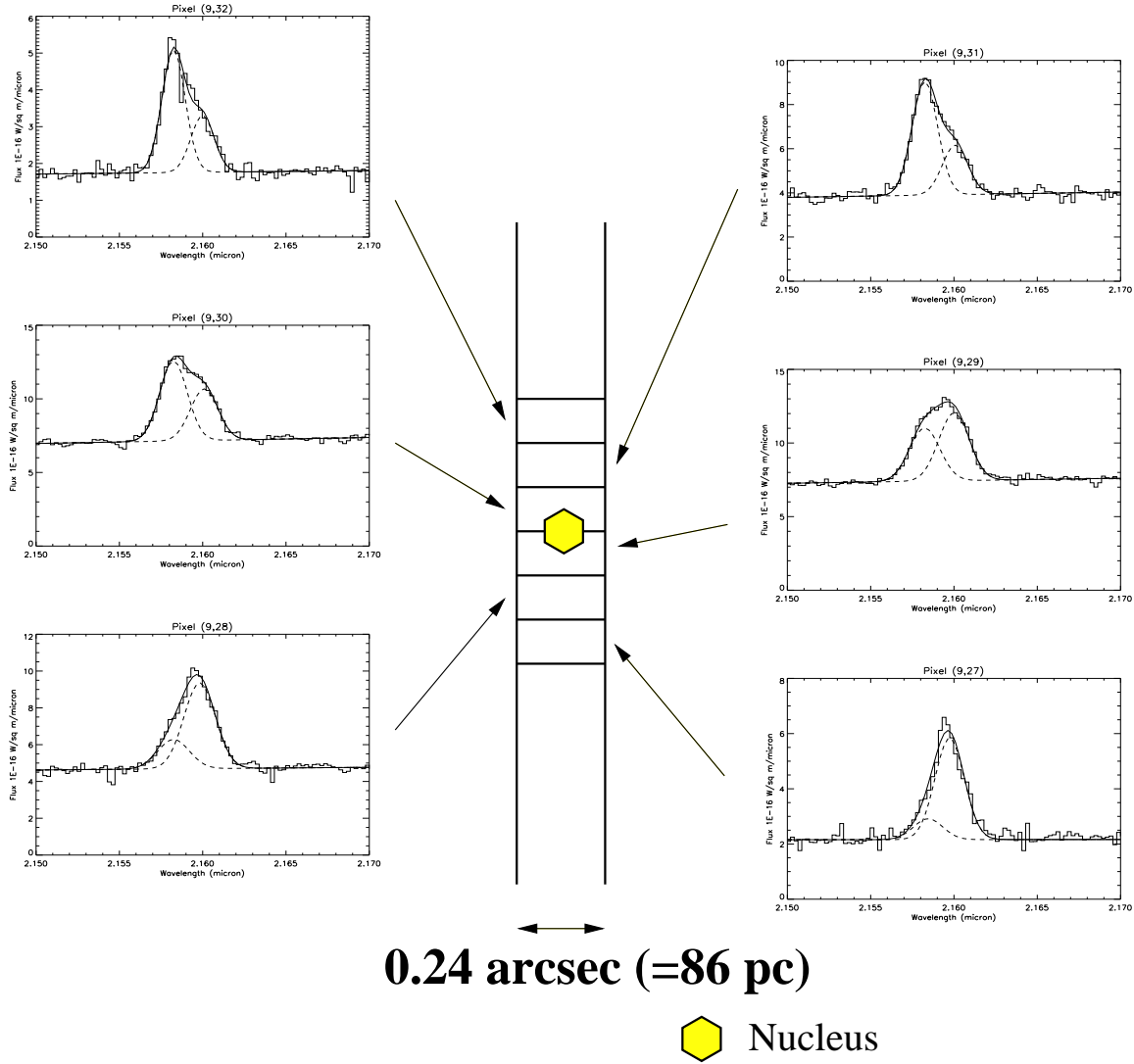}
\label{fig:S1lines}
\caption{Emission line profiles of the \H2~v=1-0~S(1) line along the peak IFU slit in the short-K data, which passes east-west through the nucleus, starting from pixel (9,27) in the east to (9,32) in the west. Adjacent pixels are separated by 0.12\arcsec~and the nucleus itself is assumed to lie mid-way between pixels (9,29) and (9,30) where the continuum peaks. Fits to the line with double gaussian profiles are also shown, and note the sharp shift in the velocity of the line peak across the nucleus.}
\label{fig:S1lines}
\end{figure*}

\section{CONCLUSIONS}
For the first time, we have spatially and kinematically resolved the hot \H2~in the circumnuclear region of NGC~1275 and found
that it could be concentrated in a rotationally-supported structure in a narrow range of radius around 50\pc~from the nucleus. A smaller amount of extended \H2~emission 
exists within a few hundred \pc~of the nucleus. X-ray heating by the active nucleus has been demonstrated to be a viable heating mechanism for the bulk of the 
thermally excited \H2~emission, and also explains why the emission is dominated by gas within a very narrow range of radii. 
Thanks to the fine pixel scale of the UIST IFU and excellent seeing, we were able to observe a sharp shift in the velocity of the
\H2~v=1-0~S(1) across the nucleus and to make a simple dynamical measurement of the black hole mass, of $3.4 \times 10^{8}$\Msun. 
The principal uncertainty in the mass arises from the uncertainty in the true inclination of the molecular gas disk to the plane of 
the sky. Nevertheless, the measured value is within 20 per cent of that inferred by Bettoni et al.~(2003) from the $M(BH)-\sigma$ relation.

The \H2~structure is likely to be an extension to smaller spatial scales of the coaxial 1.2\kpc-radius ring of CO 
emission found by Inoue et al.~(1996), which is itself the terminus of a 10\kpc-long plume of CO. This suggests that we may be directly witnessing the delivery of fuel from the galactic scale to the AGN itself. As noted in section 1, increasing numbers of cooling flow clusters appear to have large masses of CO localised in the central few tens of \kpc~and smaller masses of hot \H2~on even more compact scales. Regardless of the precise origin of the cold gas (be it from the cooling flow itself, or an interacting galaxy), the CO-\H2~connection found here for NGC 1275 may be universally important for fuelling AGN in the central galaxies of cooling flows, and hence for the regulation of heating and cooling in such cluster cores.  

 These results demonstrate the utility of the UIST IFU for high spatial and spectral resolution work of this kind, including black hole mass measurements, and promise much for the advent of similar instruments on 8-metre telescopes with adaptive optics. When coupled with sensitive CO interferometry (e.g. from ALMA) it should be possible to extend these detailed studies to more distant clusters.

\section*{ACKNOWLEDGMENTS}
UKIRT is operated by the Joint Astronomy Centre on behalf of the United Kingdom Particle Physics and Astronomy Research Council. RJW and ACE thank PPARC and the Royal Society, respectively, for support. We thank the anonymous referee for some constructive feedback.

{}

\end{document}